\documentclass[10pt,a4paper,3p,twocolumn]{elsarticle}

\usepackage[english]{babel}

\usepackage{amsfonts}
\usepackage{amssymb}
\usepackage{amsmath}
\usepackage{braket}
\usepackage{physconst}
\usepackage{lmodern}
\usepackage{color,colortbl}
\usepackage{geometry}
\usepackage{graphicx}
\usepackage{makecell}

\newcommand{\cblack}{\color[rgb]{0,0,0}}

\graphicspath{ {./images/} }

\begin{document}
\title{Robustness of a universal gate set implementation in transmon systems
via Chopped Random Basis optimal control}

\author{
	Herv\`e Ats\`e Corti\textsuperscript{1}, Leonardo Banchi\textsuperscript{2, 3}, Alessandro Cidronali\textsuperscript{4}\\
	\textit{
		\textsuperscript{1}Department of Information Engineering, University of Pisa, via G. Caruso 16, I-56122 Pisa (PI), Italy\\
		\textsuperscript{2}Department of Physics and Astronomy, University of Florence, via G. Sansone 1, I-50019 Sesto Fiorentino (FI), Italy\\
		\textsuperscript{3}INFN Sezione di Firenze, via G.Sansone 1, I-50019 Sesto Fiorentino (FI), Italy\\
		\textsuperscript{4}Department of Information Engineering, University of Florence, via di S. Marta 3, I-50139 Firenze (FI), Italy
	}
}

\begin{abstract}
	We numerically study the implementation of a universal two-qubit gate set,
	composed of CNOT, Hadamard, phase and $\pi/8$ gates, for transmon-based systems. 
	The control signals to implement such gates are obtained using  the Chopped Random Basis optimal control technique,
	with a target gate infidelity of $10^{-2}$.
	During the optimization processes we account for the leakage toward non-computational states, an important non-ideality affecting transmon qubits.
	We also test and benchmark the optimal control solutions against the introduction of Gaussian white noise and spectral distortion, two key non-idealities that affect the control signals in transmon systems.\\
	\\
	\textit{Keywords:} Transmon qubit; Qubit optimal control; Quantum gate implementation; Superconducting quantum computing; Chopped Random Basis
\end{abstract}
\maketitle

\cblack

\section{Introduction}\label{s:intro}

The engineering of a scalable and dependable quantum computational system \cite{feynman2018simulating} is strongly tied to the reliable implementation of a universal set of quantum gates \cite{nielsen2002quantum,brif2010control,glaser2015training,d2007introduction}. This task should be fulfilled with the highest possible fidelity, regardless of the number of qubits and their architecture \cite{o2009photonic}. To achieve this goal, it is mandatory to properly master the control of the underling hardware. Due to extreme operating conditions, non-idealities and constraints imposed on control signals by experimental setups, the control of quantum hardware represents one of the most critical topics in the effective development of quantum computational platforms.

Optimal control theory \cite{brif2010control,glaser2015training,d2007introduction,jurdjevic1997geometric} has had a fundamental role in advancing the techniques used to control this kind of systems. The improvement on systems control relies on the ability to manipulate the quantum hardware with optimized control signals. Among optimal control techniques, the Chopped RAndom Basis (CRAB) optimization method \cite{caneva2011chopped,rach2015dressing,muller2021one} was found to be particularly effective. CRAB operates a gradient-free minimization of system's cost functional, a compelling feature when gradient calculations are prohibitive. Although gradients of quantum evolution can be explicitly computed in small dimensional systems \cite{khaneja2005optimal}, or directly estimated via measurements on multi-qubit systems~\cite{banchi2021measuring}, an efficiently computable formula for higher dimensional multi-level systems is still missing. 
The gradient-free approach also has the beneficial side-effect to make the algorithm computationally lighter, with respect to gradient-based counterparts. Furthermore, CRAB and specially its dressed version dCRAB (dressed CRAB) \cite{rach2015dressing} are suited for optimization problems where constraints are imposed on control signals since they can easily be considered without adding extra complications to the algorithm.

Superconducting quantum systems are one of the most promising technologies that could be used to build a scalable quantum computer \cite{devoret2013superconducting,wendin2017quantum,krantz2019quantum}, and the application of optimal control techniques has shown remarkable benefits in steering these systems toward the desired evolution  \cite{rebentrost2009optimal,sporl2007optimal,werninghaus2021leakage}.
In this letter we study the performance of the CRAB optimization method \cite{muller2021one,watts2015optimizing}
to obtain reliable control signals suitable to implement a universal gate set in transmon
systems \cite{koch2007charge,bardin2020quantum,krantz2019quantum} under realistic 
imperfections, such as the qubit leakage towards non-computational energy levels.
%Gate implementation in transmon systems has been studied in literature with insightful results Consequently, we found that the application of a CRAB-based optimal control on transmon systems was an interesting topic to study.
More specifically, we test the algorithm on the CNOT (control-NOT), Hadamard, phase and $\pi/8$ gates, and benchmark
the proposed robustness \cite{lloyd2014information,kallush2014quantum,muller2020information} 
of the solutions. 
Other optimization methods have been considered in the literature for controlling transmon systems
\cite{chow2012universal,long2021universal,zahedinejad2015high}, with insightful results, 
yet the performance of CRAB-based algorithms was an open question. 
Moreover, we
test optimal control solutions against the introduction of Gaussian white noise and spectral distortion, which was 
not considered in the previous literature. The aforementioned disturbances approximate
the degrade of optimal control solutions caused by non-idealities in the control electronics. Our results should be considered as a proof of concept demonstration of the gate implementation performances, for transmon systems controlled by signals defined via CRAB-based optimization methods. In order to master the control of an experimental quantum system, it would be beneficial to directly measure in-hardware the target quantity of the optimization (e.g.~the infidelity). Such measurements, if performed carefully within an acceptable time frame, can then be used as feedback for closed-loop optimization techniques (e.g. RedCRAB \cite{muller2021one}).

This letter is organized as follows. 
In Sec.~\ref{s:control} we introduce the mathematical formulation to model the
controlled evolution of transmon systems. 
In Sec.~\ref{s:setup} we describe how we set up the optimization process. 
%In Sec.~\ref{s:numerics} we present some examples of optimal control signals to
%implement different gates, and test the resilience of the optimal control solutions 
%against noise and spectral distortion. 
In Sec.~\ref{s:numerics} we present an example of optimal control solution that
implements a CNOT gate, and test the resilience of the optimal control solutions 
against noise and spectral distortion. 
Finally, conclusions are drawn in Sec.~\ref{s:conclu}.

\section{Control of Transmon Qubits}\label{s:control}

The Hamiltonian of a quantum system can be conceptually divided in two parts, the drift Hamiltonian and the control Hamiltonian:
\begin{equation}
\hat{H}(t) = \hat{H}^{\rm drift} + \sum_{j=1}^{N} \Gamma_{j}(t) \hat{H}_{j}^{\rm control} \enspace ,
\end{equation}
where $\Gamma_{j}(t)$ are the time-dependent control signals, and $N$ is the number of control parameters.

Optimal control algorithms find the control signal profiles $\Gamma_{j}(t)$
that steer the system toward the desired evolution.  
The CRAB optimal control algorithm can be described as
follows~\cite{caneva2011chopped}: firstly, we guess an initial pulse
$\Gamma^{0}_{j}(t)$ for each control parameter; subsequently, we search for a
correction to the initial pulse in the form 
\begin{equation}
	\Gamma_{j}(t) = \Gamma^{\rm CRAB}_{j}(t) = \Gamma^{0}_{j}(t) \cdot G_{j}(t) \enspace .
	\label{CRABG}
\end{equation}
The functions $G_{j}(t)$ are expressed by a limited amount of spectral components $N_{c}$:
\begin{equation}
G_{j}(t) = \sum_{k=1}^{N_{c}} c_{j}^{k} g_{j}^{k} (\omega_{j}^{k}t) \enspace ,
\end{equation}
where the terms $g_{j}^{k}$ represent some basis functions for a Fourier-type series expansion, $\omega_{j}^{k}$ represent angular frequencies, and $c_{j}^{k}$ correspond to the coefficients associated to each basis function \cite{caneva2011chopped}. In the CRAB implementation adopted by Qutip Python library \cite{johansson2012qutip}, the functions $g_{j}^{k}$ correspond to sinusoidal and cosinusoidal functions. 
The core aspect of CRAB, that is, the basis frequencies randomization is performed by Qutip through the following approach:
\begin{equation}
\omega_{j}^{k} = k \cdot \frac{2\pi}{T} + r_{j}^{k} \enspace ,
\end{equation}
where $r_{j}^{k}$ are random frequency offsets uniformly distributed over $[-0.5, 0.5)$, 
and $T=\mathcal O(1)$ is the dimensionless time duration of the gate implementation (i.e. the time duration of the control signals) -- for transmon systems we may consider time in nanoseconds, so the resulting frequencies are assumed in the GHz regime. 
With $r_{j}^{k} = 0$ (i.e. without the basis frequency randomization) the terms $\omega_{j}^{k}$ are simply the Fourier harmonics.
%Consequently, if we consider the nanosecond as the order of magnitude of $T$, the order of magnitude of $\omega_{j}^{k}$ and $r_{j}^{k}$ becomes $rad \cdot GHz$.
Qutip's randomization of basis frequencies is slightly different from the randomization process proposed in the original CRAB paper \cite{caneva2011chopped} where the randomization is carried out through the following equation:
\begin{equation}
\omega_{j}^{k} = (k + r_{j}^{k}) \cdot \frac{2\pi}{T} \enspace ,
\end{equation}
in this case $r_{j}^{k}$ correspond to dimensionless random numbers. 
However, despite the differences between the two randomization processes, we expect to have similar optimization results since both are able to create a set of random basis frequencies that can be used to randomize the basis functions.
The randomization of basis functions greatly improves the algorithm convergence, despite causing the loss of the orthonormality condition.
CRAB optimization problem consists in finding the optimal coefficients $c_{j}^{k}$ that minimize a suitable cost functional, such as the gate infidelity, as we describe in more detail hereinafter. This task is then carried out via direct-search, for example via the Nelder-Mead simplex method \cite{nelder1965simplex,caneva2011chopped}.

The low-energy Hamiltonian of two-transmon system with tunable coupling (also dubbed ``g-mon'' \cite{chen2014qubit}), in the interaction picture and after the rotating wave approximation, can be written as a Bose-Hubbard type Hamiltonian 
\cite{bardin2020quantum,krantz2019quantum,niu2019universal}:
\begin{equation}
\begin{aligned}
\hat{H}_{BH} &= \sum_{j=1}^{2}\Big[\delta_{j}(t) \hat{n}_{j} + \dfrac{\eta}{2} \hat{n}_{j}(\hat{n}_{j} - 1) + \\
						 &+ i(\hat{a}_{j} e^{i\psi_{j}(t)} - \hat{a}_{j}^{\dagger} e^{-i\psi_{j}(t)}) F_{j}(t)\Big] + \\
&+ g(t)(\hat{a}_{1} \hat{a}_{2}^{\dagger} + \hat{a}_{2} \hat{a}_{1}^{\dagger}) \enspace .
\end{aligned}
\label{Ham_No_Constraints}
\end{equation}
The operators $\hat{n}_{j}$ correspond to number operators, and $\hat{a}_{j}$ ($\hat{a}_{j}^{\dagger}$) correspond to annihilation (creation) operators. The terms $\delta_{j}(t)$ can be controlled by externally modulating the magnetic flux in each transmon \cite{bardin2020quantum}, and correspond to the detuning parameters, that is how much the $j$-th qubit oscillation frequency departs from that of the control resonator. The parameter $\eta$ accounts for the anharmonicity of the qubit energy levels, with its value selected at the design stage. In our simulations we fixed $\eta=0.2$ GHz (approximately $8.3 \cdot 10^{-7}$ eV), a typical value for transmon devices \cite{bardin2020quantum,krantz2019quantum}. The anharmonicity $\eta$ can't be controlled by electromagnetic signals.
The terms $F_{j}(t)$ and $\psi_{j}(t)$ correspond to the amplitudes and phases of the microwave electromagnetic signals used to drive the $j$-th qubit state. Since the Hamiltonian is defined in a rotating frame of reference, in the laboratory frame of reference the drive control parameters $F_{j}(t)$ acts as an envelope for a sinusoidal carrier oscillating at the qubit resonant frequency \cite{bardin2020quantum,krantz2019quantum}.
Lastly the term $g(t)$ accounts for qubit-qubit coupling, and it is controllable through an external magnetic flux \cite{chen2014qubit}. 
The simulated gates require implementation times in the order of tens of nanoseconds; accordingly, the order of magnitude of the energy carried by control signals is GHz (approximately $10^{-6}$ eV).

In order to simplify both the Hamiltonian and the control signals, without loss of generality we impose:
\begin{equation}
\psi_{j}(t) = 0, \enspace \forall j .
\end{equation}
Consequently, the system Hamiltonian can be rewritten as:
\begin{equation}
\begin{aligned}
\hat{H}_{BH} &= \sum_{j=1}^{2}\Big[\delta_{j}(t) \hat{n}_{j} + \dfrac{\eta}{2} \hat{n}_{j}(\hat{n}_{j} - 1) + \\
						 &+ i(\hat{a}_{j} - \hat{a}_{j}^{\dagger}) F_{j}(t)\Big] + \\
&+ g(t)(\hat{a}_{1} \hat{a}_{2}^{\dagger} + \hat{a}_{2} \hat{a}_{1}^{\dagger}) \enspace .
\end{aligned}
\label{Ham}
\end{equation}
We show that, even with such constraints, we have been able to implement all quantum gates of the chosen universal set.
All the other terms with time dependence in \eqref{Ham} represent control parameters that we optimized through CRAB in order to implement the desired gates.

\section{Optimizations Setup}\label{s:setup}

We carried out the optimizations using Qutip's CRAB implementation to define
the control parameters of the Hamiltonian \eqref{Ham}, using  
10 sinusoidal and 10 cosinusoidal randomized basis functions for each parameter.
All our optimizations adopt $\Gamma^{0}_{j}(t) = 1$ as initial pulse guess; therefore, we can reformulate the optimization problem in \eqref{CRABG} as $\Gamma^{CRAB}_{j}(t) = G_{j}(t).$ In case an experimental setup needs control signals with a constrained envelope, the initial pulse guess can introduce the required constraints in the optimization process.

As cost functional for our optimizations \cite{rembold2020introduction} we set the gate infidelity. The mathematical definition of the adopted gate infidelity (from now on infidelity) is as follows: 
\begin{equation}
I = 1 - \dfrac{1}{d^2} | {\rm Tr}(\hat{U}_{\rm target}^{\dagger} \hat{U}(T)) | \enspace ,
\label{infidelity}
\end{equation}
where $d$ corresponds to the dimension of the Hilbert space,
where the operators $\hat{U}(T)$ and $\hat{U}_{\rm target}$ act,
$\hat U(T)$ refers to the solution of the time-dependent Schr\"odinger equation up to time $T$, with Hamiltonian \eqref{Ham}, and 
$\hat U_{\rm target}$ is the target gate.
We fixed the gate implementation time at $T=40$ ns, a value above the quantum speed limit for this type of systems \cite{taddei2013quantum,margolus1998maximum,bhattacharyya1983quantum}, and in line with state-of-the-art implementation times \cite{barends2014superconducting}. The infidelity \eqref{infidelity} quantified how much the gate applied to our system via control pulses deviated from the ideal gate under investigation. 

Non-computational energy levels of transmon qubits represent an important non-ideality in experimental gate implementations \cite{zahedinejad2015high}. These extra levels can be exploited to implement fast and reliable gates \cite{krantz2019quantum}, yet we need to avoid unwanted qubit leakages from computational to non-computational energy levels. 
We account for this non-ideality by approximating an hardware qubit via a three-level quantum system, interacting via a truncated version of the Hamiltonian \eqref{Ham}. We simulated a three-level subspace since for transmon qubits the third energy level represents the most dominant contribution to leakages outside the computational subspace defined by the two lowest energy levels \cite{rebentrost2009optimal}. Moreover, simulating qubits with more than three energy levels significantly increases the simulation cost in Qutip, leading to unpractical optimization times. 
In our approximation, the first and second energy levels correspond to the qubit computational states $\ket 0$ and $\ket 1$, respectively, while the third level corresponds to the non-computational state $\ket 2$ \cite{rebentrost2009optimal,werninghaus2021leakage}. All gates matrices have been reshaped in order to act on this three-level subspace.
We put no constraints on how $\hat U(T)$ acts on the third energy level since the population of this non-computation state is an imperfection itself. Accordingly, to calculate the infidelity, we firstly discard the non-computational energy levels from $\hat U(T)$, and then compute Eq.~\eqref{infidelity} with the ideal two-level target gate $U_{\rm target}$. 
This approach has the advantage of reducing the numerical complexity of the optimization.

\section{Optimal Control Solutions Analysis}\label{s:numerics}

In this section we present the solutions of the different numerical experiments 
performed with the CRAB algorithm. 
We considered a gate successfully implemented when its infidelity was minimized below the threshold, set at $10^{-2}$. We decided to adopt this threshold value since it approximates the infidelity magnitude of the best performing gates implemented on experimental transmon systems \cite{arute2019quantum,kelly2014optimal,kandala2021demonstration}. Furthermore, we gathered 30 different optimal control solutions for each gate of the chosen universal set in order to benchmark the average and best performances, as well as to weigh lucky gate implementations.
All optimization attempts have been able to define the control solutions at the first try with our target infidelity of $10^{-2}$. The average and the minimum infidelities that we obtained are shown in Table~\ref{table:inf}. %We considered a gate implementation ideal when the optimal control signals that implement the target gate are not subjected to the introduction of noise or spectral distortion. The average and the minimum infidelities that we obtained for the ideal gate implementations are shown in Table \ref{table:inf}.

\begin{table}[]
    \begin{tabular}{|c|c|c|c|c|}

    \hline
    {\makecell{}}
    & {\makecell{Average\\ Infidelity}}
    & {\makecell{Minimum\\ Infidelity}} \\ \hline

    CNOT     & $9.9930 \cdot 10^{-3}$ & $9.9577 \cdot 10^{-3}$ \\ \hline
    Hadamard & $9.9944 \cdot 10^{-3}$ & $9.9742 \cdot 10^{-3}$ \\ \hline
    phase    & $9.9922 \cdot 10^{-3}$ & $9.9513 \cdot 10^{-3}$ \\ \hline
    $\pi/8$  & $9.9938 \cdot 10^{-3}$ & $9.9290 \cdot 10^{-3}$ \\ \hline
      
    \end{tabular}
    \caption{Average and minimum infidelities with 30 different optimal control solutions. }
	%gate implementations for each gate of the universal set.}
    \label{table:inf}
\end{table}

\begin{figure}[t]
\includegraphics[scale=0.45]{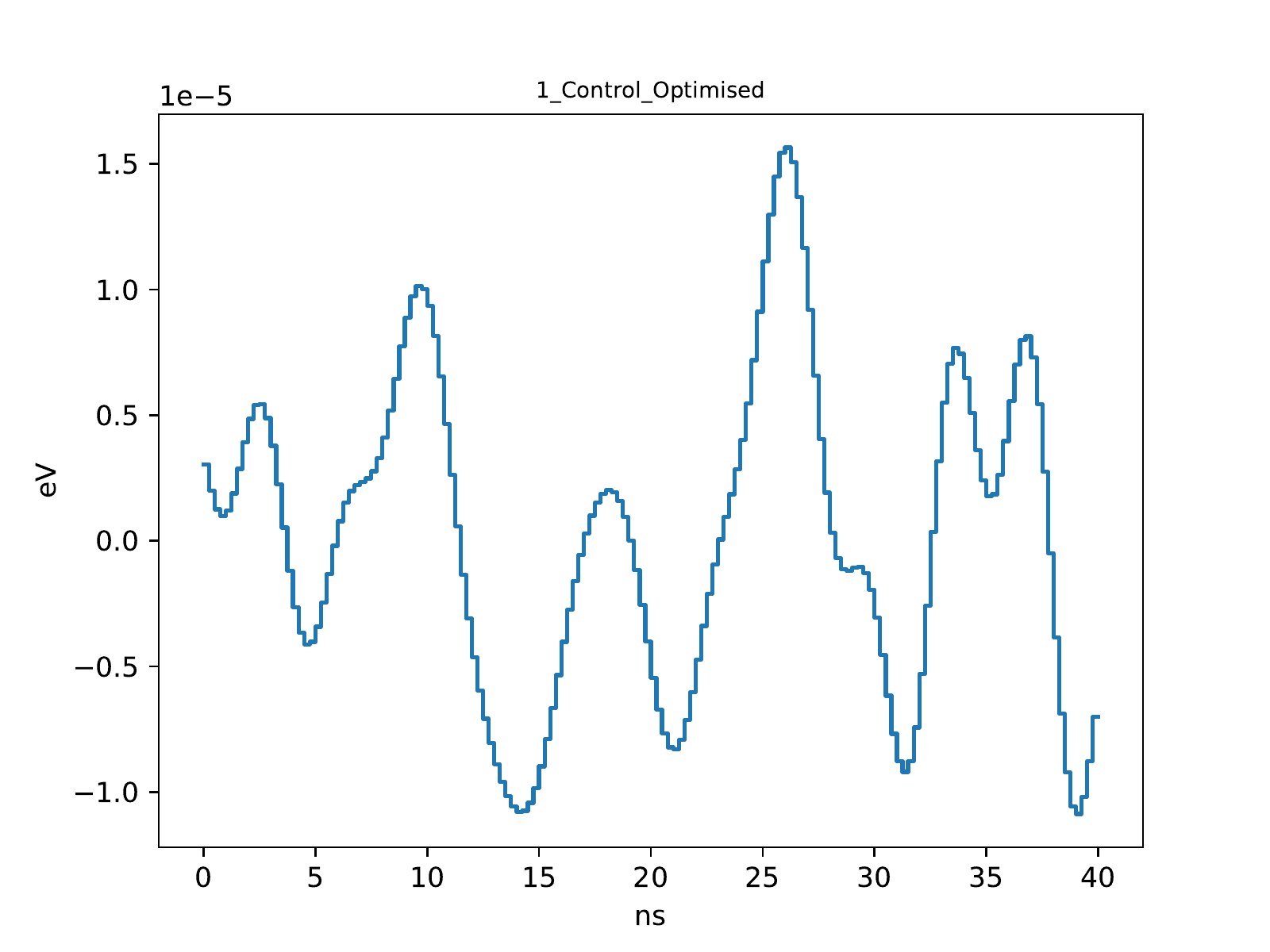}
\caption{Amplitude $F_2(t)$ to implement a CNOT gate via the transmon Hamiltonian Eq.~\eqref{Ham}. }
\label{fig:control}
\end{figure}

\begin{figure}[t]
\includegraphics[scale=0.45]{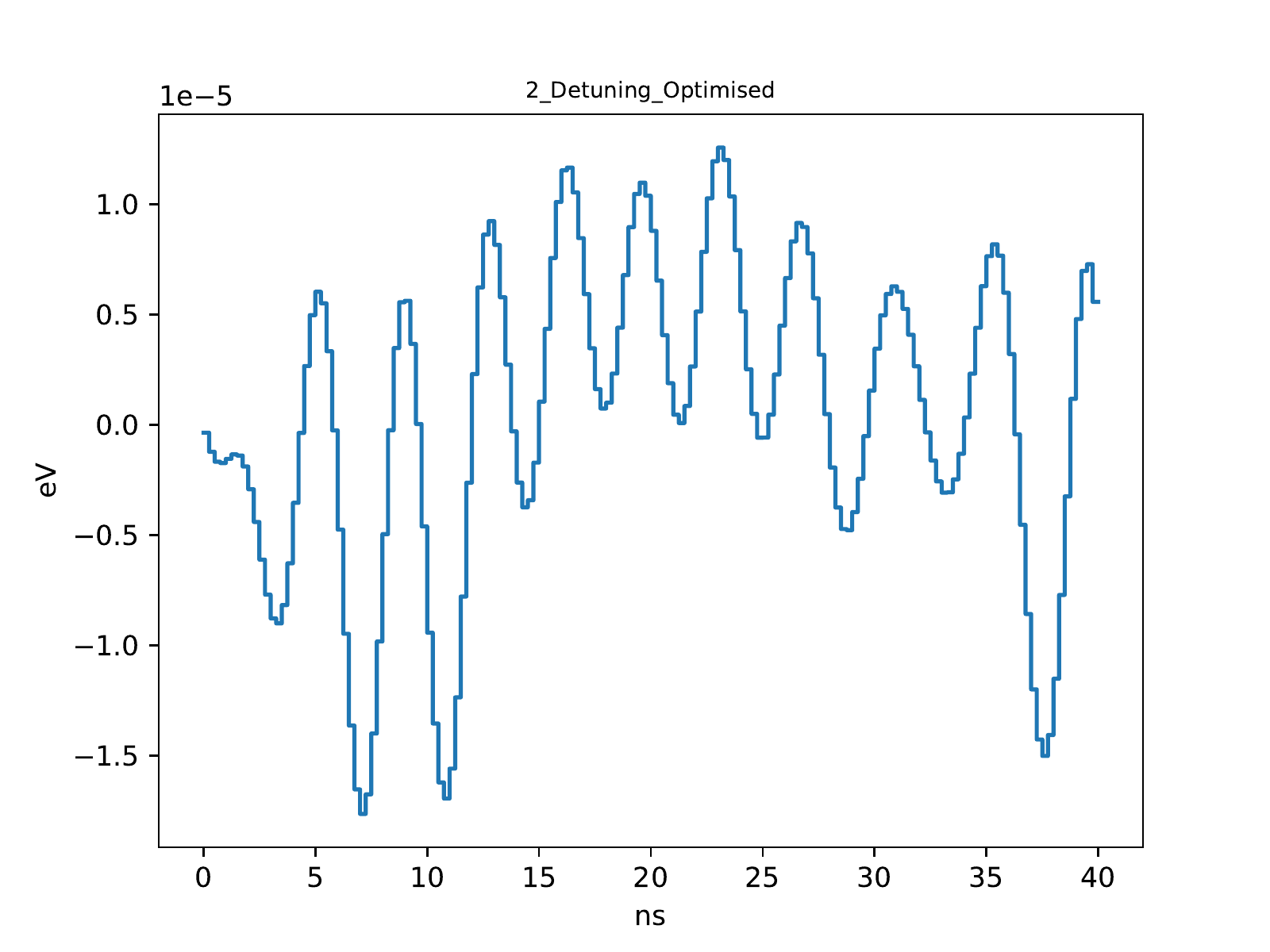}
\caption{Detuning $\delta_1(t)$ to implement a CNOT gate via the transmon Hamiltonian Eq.~\eqref{Ham}. }
\label{fig:detuning}
\end{figure}

\begin{figure}[t]
\includegraphics[scale=0.45]{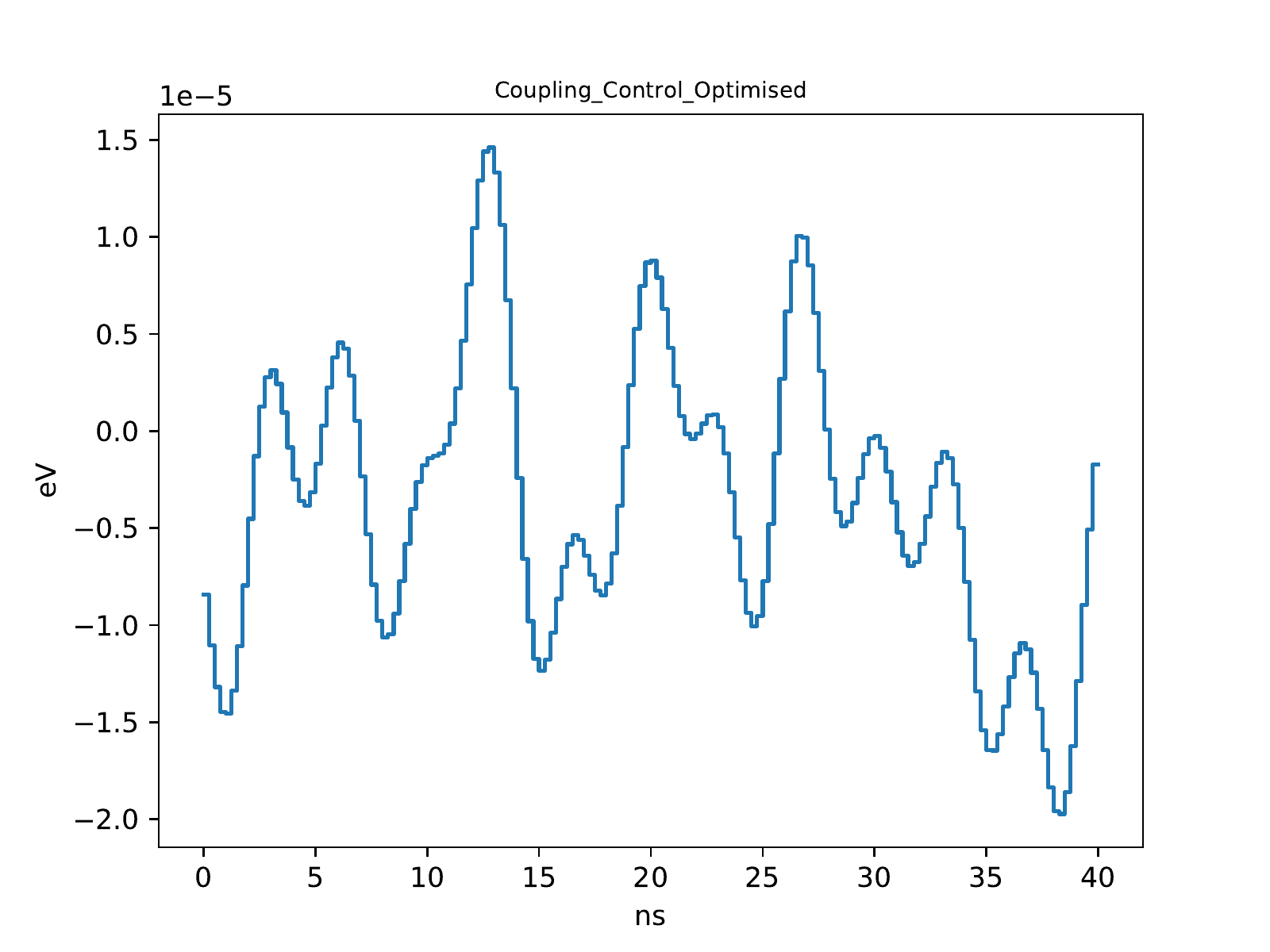}
\caption{Coupling $g(t)$ to implement a CNOT gate via the transmon Hamiltonian Eq.~\eqref{Ham}. }
\label{fig:coupling}
\end{figure}

The control signals that compose one of the 30 optimal control solutions that implement the CNOT gate are shown in Figures \ref{fig:control}, \ref{fig:detuning} and \ref{fig:coupling}. This control solution has an infidelity of $9.9982 * 10^{-3}$.
Fig.~\ref{fig:control} shows the profile of the amplitude $F_j(t)$ from Eq.~\eqref{Ham} on the target qubit $j=2$. Fig.~\ref{fig:detuning} shows the profile of the control qubit detuning $\delta_{1}(t)$, while  Fig.~\ref{fig:coupling} shows the profile of the qubit-qubit coupling $g(t)$.
Figs.~4, 5 and 6 show the spectra obtained by FFT analysis of the optimized control signals showed in Figure 1, 2 and 3, respectively.

\begin{figure}[t]
\includegraphics[scale=0.45]{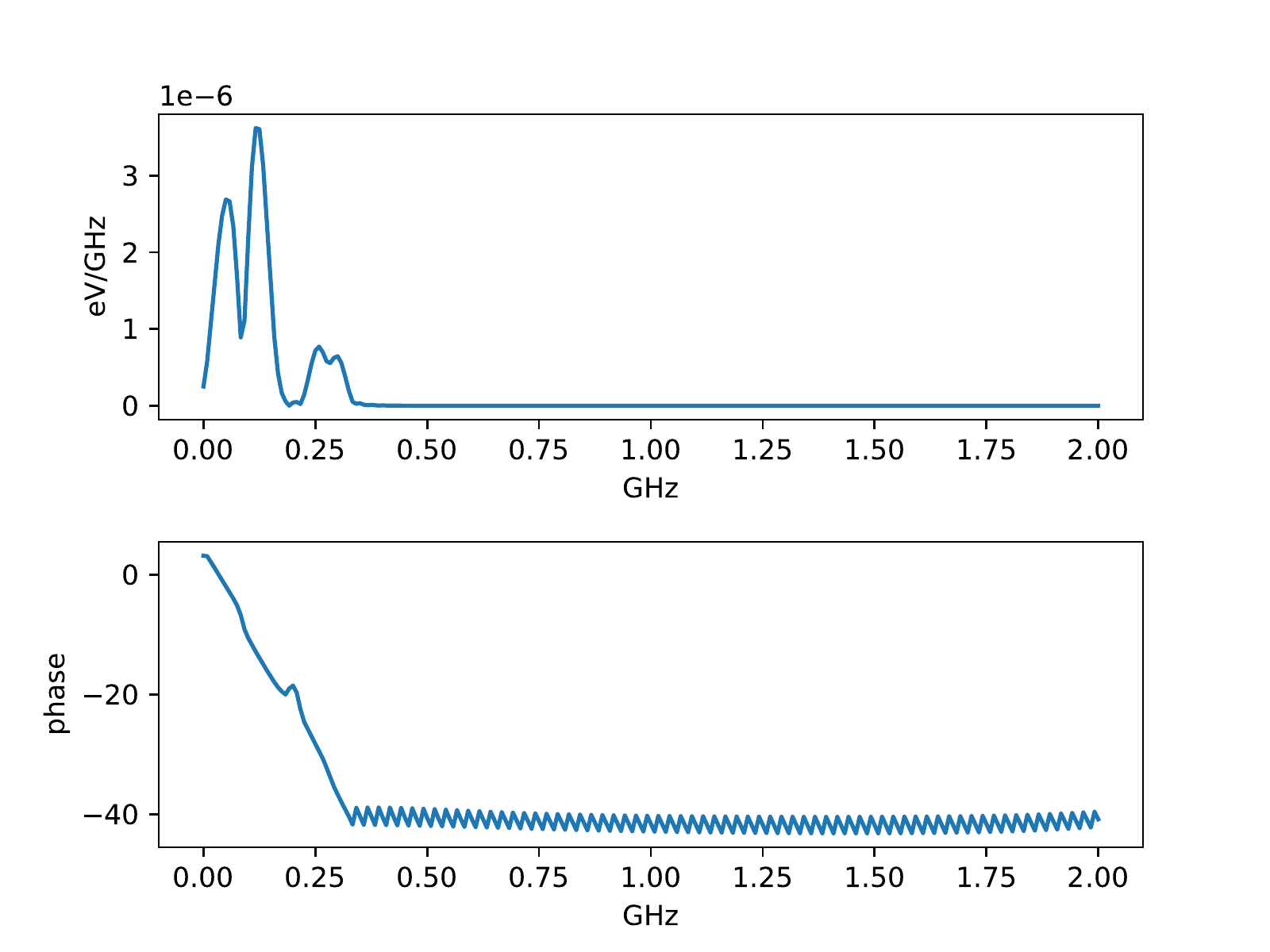}
\caption{Fourier spectrum of the control signal $F_2(t)$ shown in Fig.~1.}
\end{figure}
\begin{figure}[t]
\includegraphics[scale=0.45]{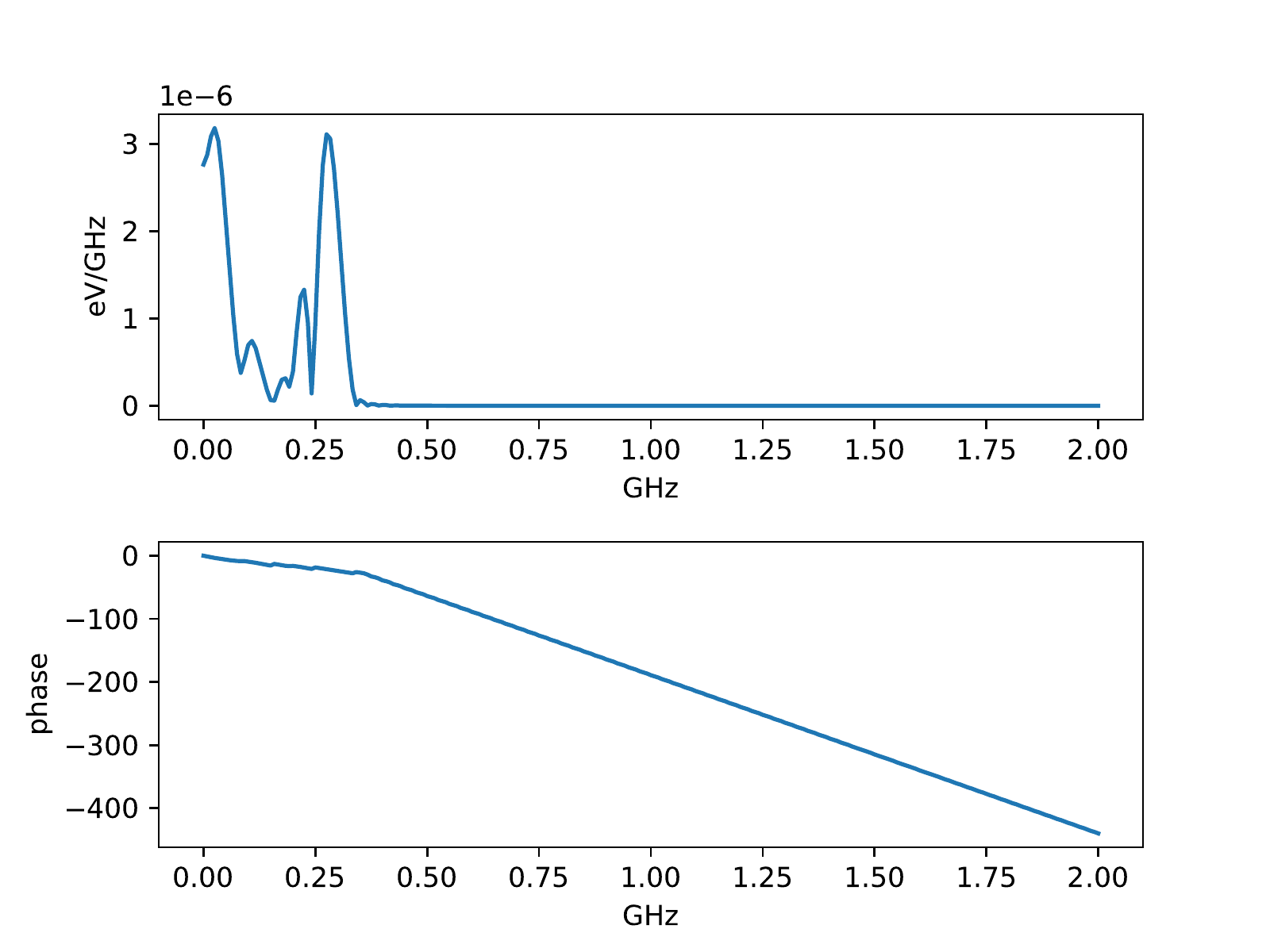}
\caption{Fourier spectrum of the control signal $\delta_1(t)$ shown in Fig.~2.}
\end{figure}
\begin{figure}[t]
\includegraphics[scale=0.45]{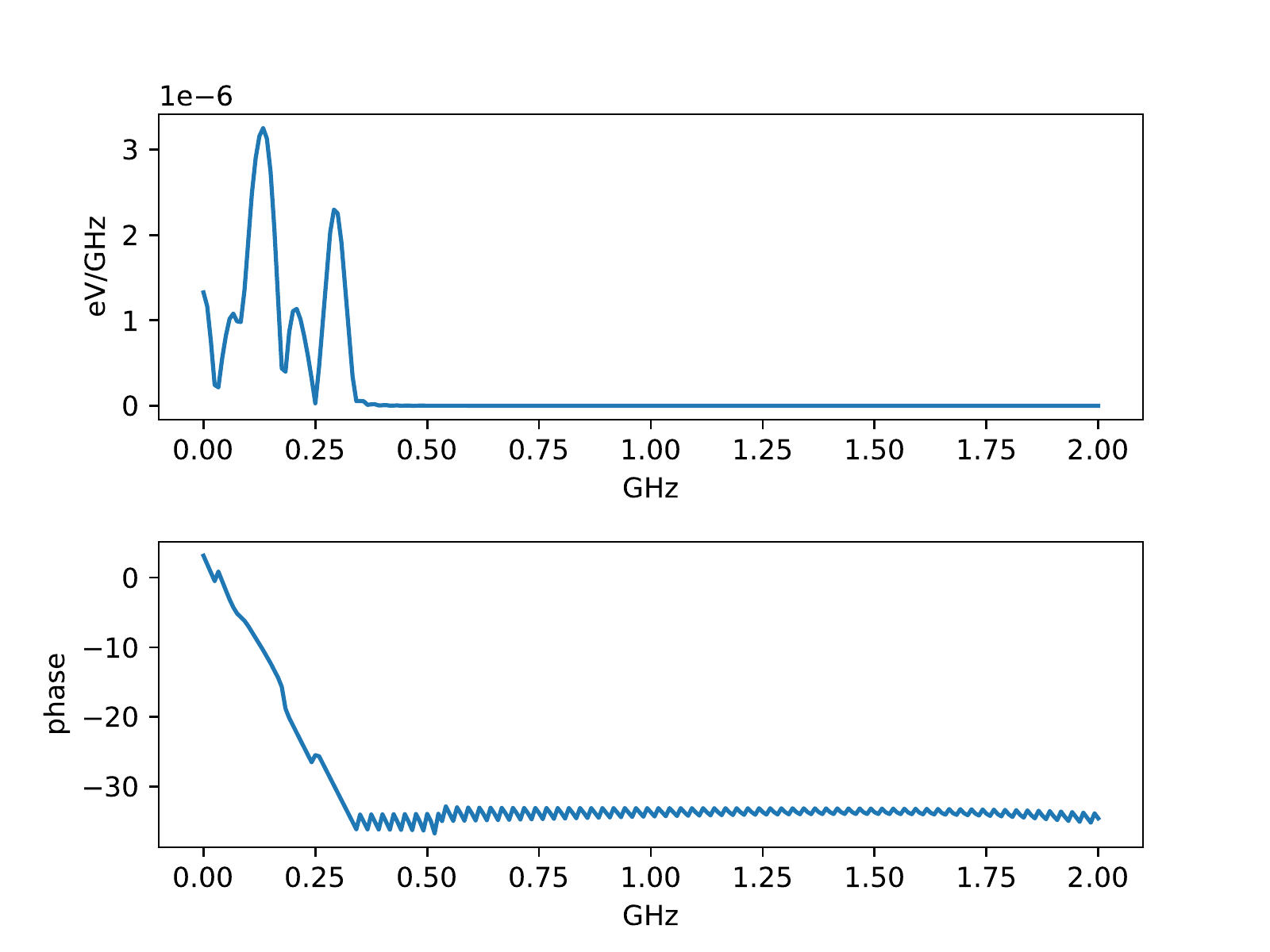}
\caption{Fourier spectrum of the control signal $g(t)$ shown in Fig.~3.}
\end{figure}

In our theoretical approach we considered some of the main sources of  
problems and non-idealities that arise in experimental setups. In particular, we simulated the noise carried by control signals and the distortion that affects their spectra. The noise that we introduced on control signals is Gaussian and white, and can be considered as a simple approximation of stochastic noise sources affecting qubits \cite{bardin2020quantum,krantz2019quantum}, such as the noise leaking from control electronics and the noise generated by the quantum hardware itself. There are also other types of noise affecting solid-state systems, and not considered in this work; in particular, $1/f$ noise sources are among the most important causes of decoherence in this kind of quantum systems \cite{paladino20141,schlor2019correlating,montangero2007robust}. 

The following equation describes the inclusion of noise in control signals:
\begin{align}
\begin{split}
	\Gamma_{j}^{\rm noise}(t) &  = \Gamma^{\rm CRAB}_{j}(t) + \Delta_{j}^{\rm noise}(t) \\
														&= G_{j}(t) + \Delta_{j}^{\rm noise}(t) \enspace .
\end{split}
\end{align}
The parameters $\Delta_{j}^{\rm noise}(t)$ correspond to a time dependent white noise. To evaluate the noise resilience of existing optimal control solutions, we superimposed $\Delta_{j}^{\rm noise}(t)$ to the optimized control signals \cite{kallush2014quantum}. We considered the noise tolerable by a control solution when 30 different noise realizations $\Delta_{j}^{\rm noise}(t)$ with a guessed standard deviation $\sigma_{\Delta^{noise}}$ were able to give 30 consecutive gate implementations with an infidelity below $10^{-2}$. In order to define $\sigma_{\Delta^{noise}}$, our algorithm proceeded by reducing it with $-1$ dB steps starting from $0.1$ GHz (approximately $4.1 \cdot 10^{-7}$ eV), until the noise was tolerable by the tested control solution. We repeated this test for each of the 30 optimal control solutions that implement each of the gates of the universal set.
Following the above definition, we call $\sigma_{\Delta^{noise}}^{\rm tol.}$ as the maximum tolerated value of $\sigma_{\Delta^{noise}}$. 
The $\sigma_{\Delta^{noise}}^{\rm tol.}$ of all the simulated gates are shown in Table \ref{table:noise}.
We found that,
counter-intuitively, the introduction of noise for some given standard deviation levels improved the optimization performances to some extent.
This finding is at presently outside the scope of this work, although a similar behavior was observed with parametric quantum circuits \cite{gentini2020noise}; it will be further investigated in the future.
Fig.~7 shows how $\sigma_{\Delta^{noise}}$ affects the infidelity of a CNOT gate, implemented through the optimized control signals shown in Figure 1, 2 and 3.

\begin{figure}[t]
\includegraphics[scale=0.45]{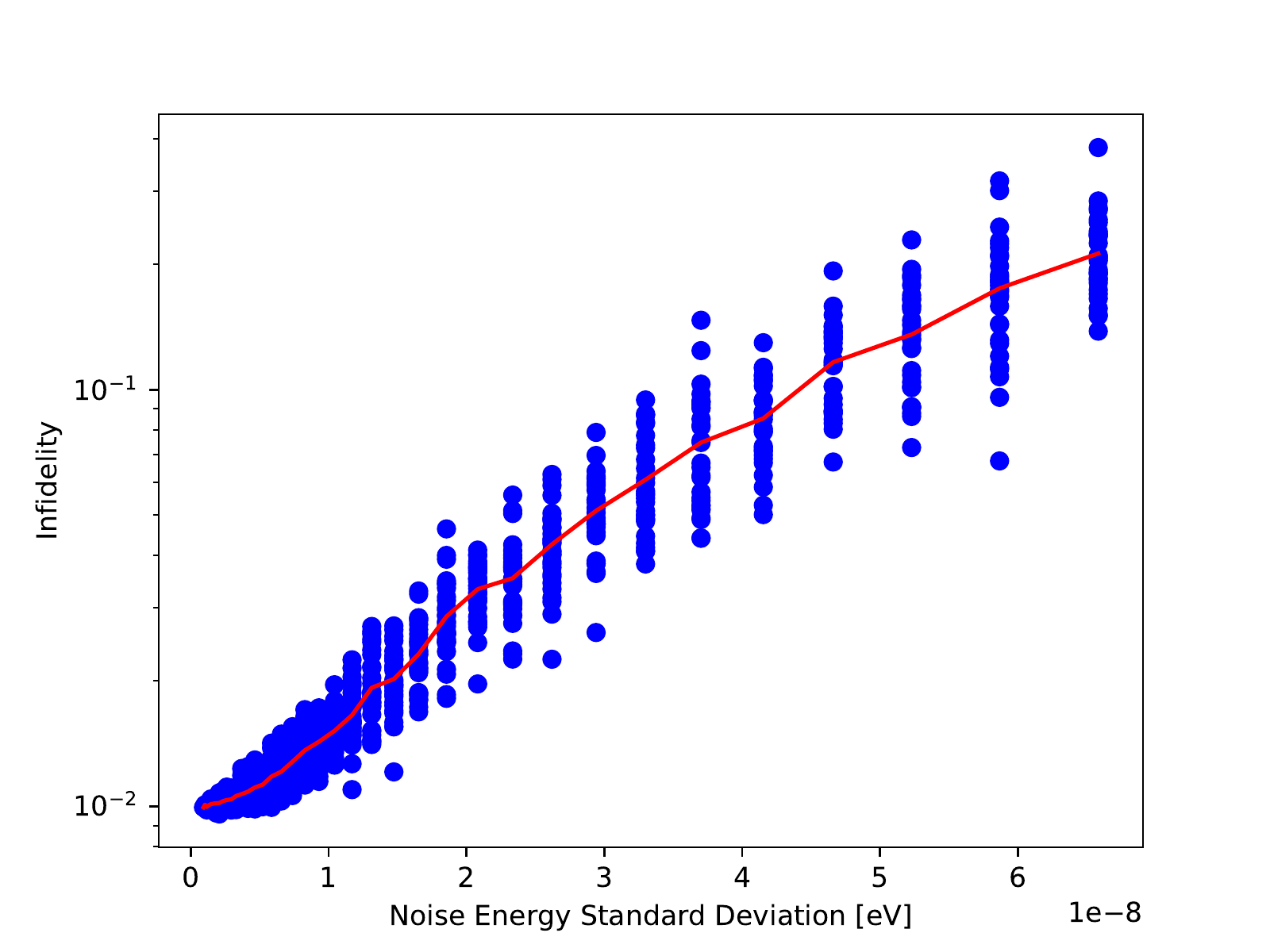}
\caption{White noise standard deviation  $\sigma_{\Delta^{noise}}$ versus gate infidelity of a CNOT gate implementation.}
\end{figure}

\begin{table}[]
    \begin{tabular}{|c|c|c|c|c|}

    \hline
    {\makecell{}}
    & {\makecell{Average\\ White Noise}}
    & {\makecell{Maximum\\ White Noise}} \\ \hline

    CNOT     & $1.52 \cdot 10^{-8}$ eV & $1.93 \cdot 10^{-8}$ eV \\ \hline
    Hadamard & $1.55 \cdot 10^{-8}$ eV & $2.38 \cdot 10^{-8}$ eV \\ \hline
    phase    & $1.54 \cdot 10^{-8}$ eV & $1.96 \cdot 10^{-8}$ eV \\ \hline
    $\pi/8$  & $1.54 \cdot 10^{-8}$ eV & $2.10 \cdot 10^{-8}$ eV \\ \hline
      
    \end{tabular}
    \caption{White noise standard deviations $\sigma_{\Delta^{noise}}$ tolerated 
		by the system when simulating different universal gates.}
	\label{table:noise}
\end{table}

In order to account for the problems caused by control signal distortions, we have also perturbed the basis coefficients of the optimal control solutions. Such perturbation can be considered as an approximation of systematic noise sources affecting control signals \cite{bardin2020quantum,krantz2019quantum}, for example, the spectral distortion introduced by digital to analog conversions, transport mediums (e.g. wires and striplines) and undesired couplings with the quantum chip physical structure.
The following equation describes the inclusion of the perturbation in the basis coefficients of an optimal control solution:
\begin{align}
\begin{split}
	\Gamma^{\rm dist.}_{j}(t) &= \Gamma^{\rm CRAB}_{j}(t) + \Delta^{\rm dist.}_{j}(t) \\
&=\Gamma^{\rm CRAB}_{j}(t) + \sum_{k=1}^{N_{c}} \Delta_{j}^{k} g_{j}^{k} (\omega_{j}^{k}t)  \\
&=G_{j}(t) + \sum_{k=1}^{N_{c}} \Delta_{j}^{k} g_{j}^{k} (\omega_{j}^{k}t)  \\
&=\sum_{k=1}^{N_{c}} (c_{j}^{k} + \Delta_{j}^{k}) g_{j}^{k} (\omega_{j}^{k}t) \enspace .
\end{split}
\end{align}

The parameters $\Delta_{j}^{k}$ superimpose a random perturbation, on the $j$-th coefficient $c_{j}^{k}$ of an optimized control signal. To evaluate the resilience of the optimal control solutions against distortion, we superimpose $\Delta_{j}^{k}$ to the coefficients of basis functions of its control signals \cite{kallush2014quantum}. We considered the spectral distortion tolerable by a control solution when 30 different realizations of a random distortion $\Delta^{\rm dist.}_{j}(t)$ with a guessed standard deviation $\sigma_{\Delta}$ for its coefficients $\Delta_{j}^{k}$ were able to give 30 consecutive gate implementations with an infidelity below $10^{-2}$. In order to define $\sigma_{\Delta}$, our algorithm proceeded by reducing it with $-1$ dB steps, starting from $0.1$ GHz (approximately $4.1 \cdot 10^{-7}$ eV), until the disturbance was tolerable by the tested control solution. 
We repeated this test for each of the 30 optimal control solutions that implement each of the gates of the universal set.
Following the above definition, we call $\sigma_\Delta^{\rm tol.}$ as the maximum tolerated value of $\sigma_\Delta$. The $\sigma_\Delta^{\rm tol.}$ of all the simulated gates are shown in Table \ref{table:dist}. 
Even during this test we observed that some levels of perturbation of basis coefficients slightly improved the optimization performances.
Fig.~8 shows how $\sigma_{\Delta}$ affects the infidelity of a CNOT gate implemented through the optimized control signals shown in Figure 1, 2 and 3. 

\begin{figure}[t]
\includegraphics[scale=0.45]{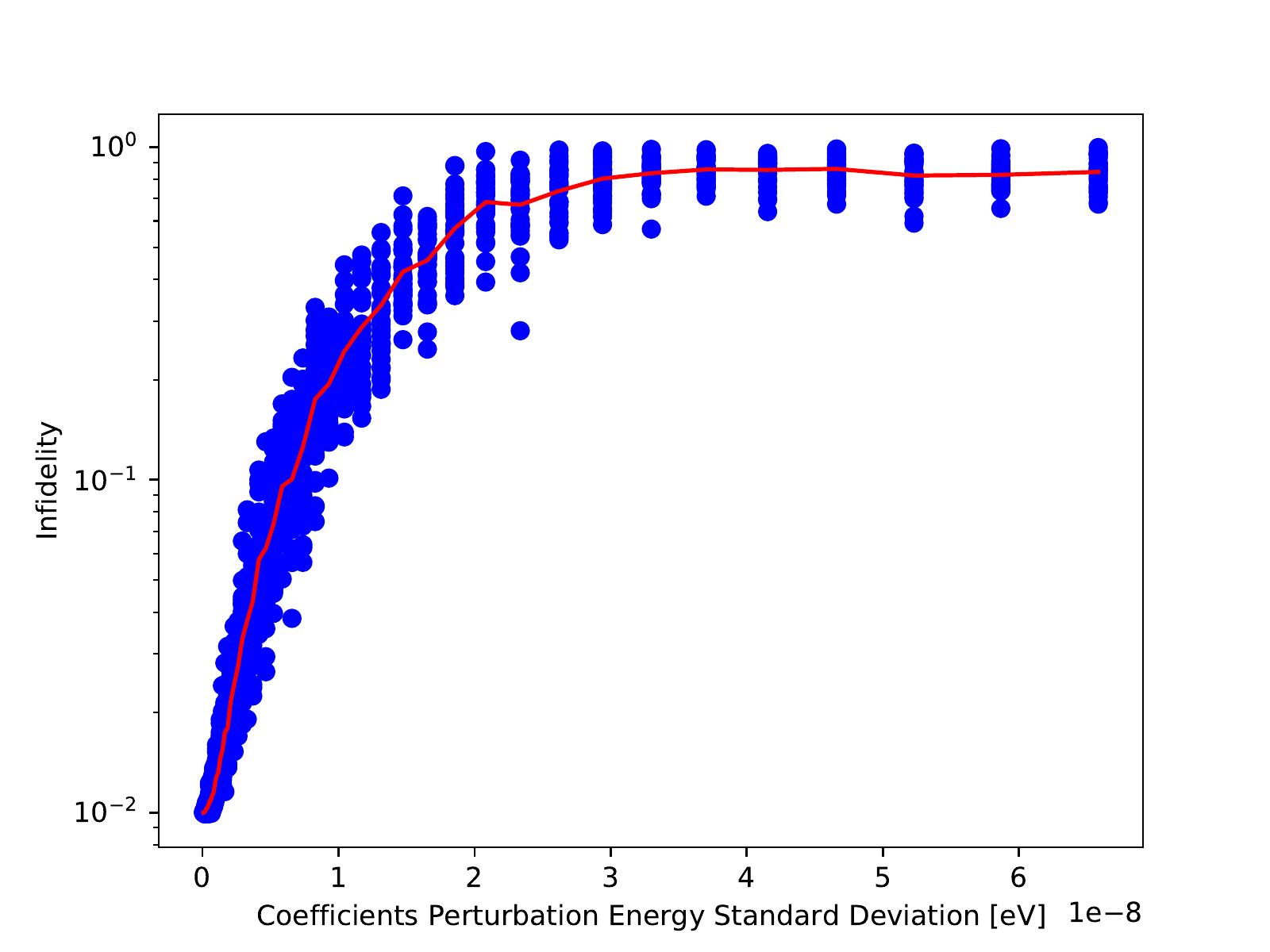}
\caption{Perturbation of a coefficient standard deviation (i.e. $\sigma_{\Delta}$) versus gate infidelity of a CNOT gate implementation.}
\end{figure}

\begin{table}[]
    \begin{tabular}{|c|c|c|c|c|}

    \hline
    {\makecell{}}
    & {\makecell{Average\\ Coeff. Pert.}}
    & {\makecell{Maximum\\ Coeff. Pert.}} \\ \hline

    CNOT     & $1.09 \cdot 10^{-9}$ eV & $1.87 \cdot 10^{-9}$ eV \\ \hline
    Hadamard & $1.00 \cdot 10^{-9}$ eV & $2.01 \cdot 10^{-9}$ eV \\ \hline
    phase    & $1.11 \cdot 10^{-9}$ eV & $1.76 \cdot 10^{-9}$ eV \\ \hline
    $\pi/8$  & $1.03 \cdot 10^{-9}$ eV & $1.83 \cdot 10^{-9}$ eV \\ \hline
      
    \end{tabular}
    \caption{White noise standard deviations $\sigma_{\Delta}$ tolerated 
		by the system when simulating different universal gates.}
	\label{table:dist}
\end{table}

We now introduce a qualitative reasoning to compare how noise and perturbations on the coefficients affect the optimal control solutions. The scope of the comparison is to find if one of the two disturb sources can be better tolerated by control signals. 
Distortions in the frequency domain can be used to construct a time dependent distortion signal, mathematically defined as:
\begin{equation}
\Delta^{\rm dist.}_{j}(t) = \sum_{k=1}^{N_{c}} \Delta_{j}^{k} g_{j}^{k} (\omega_{j}^{k}t) \enspace .
\end{equation}
All coefficient perturbations $\Delta_{j}^{1,...,N_{c}}$ have a normal distribution with 
zero expectation value and standard deviation equal to $\sigma_{\Delta}$. Since $g_{j}^{k}$ correspond to sinusoidal and cosinusoidal functions, it is possible to observe that:
\begin{align}
	\begin{split}
	\Delta^{\rm dist.}_{j}(t) &= \sum_{k=1}^{N_{c}} \Delta_{j}^{k} g_{j}^{k} (\omega_{j}^{k}t) 
									   \leq \sum_{k=1}^{N_{c}} | \Delta_{j}^{k} | \simeq N_{c} \cdot \mu \enspace ,
\end{split}
\label{d_limit}
\end{align}
where in the last approximation we have replaced the stochastic coefficient $|\Delta_j^k|$ with their mean $\mu$.
Since $|\Delta_j^k|$ are identically distributed and follow a half-normal distribution, we get 
\begin{equation}
\mu = \sigma_{\Delta} \cdot \sqrt{\dfrac{2}{\pi}} \enspace .
\label{d_abs_mean}
\end{equation}
The distortion $\Delta^{\rm dist.}_{j}(t)$ has the characteristic of a periodic source that lies in the band of control signals, and its energy is approximately contained within $\mu^{\rm dist.}=N_c \cdot \mu$. 
The average tolerated $\sigma_{\Delta}^{\rm tol.}$ that we obtained for 30 implementations of a CNOT gate is $\sigma_{\Delta}^{\rm tol.} \simeq 1.09 \cdot 10^{-9}$ eV. From Eq.\eqref{d_abs_mean} we obtain $\mu \simeq 0.87 \cdot 10^{-9}$ eV for a single perturbation of a basis coefficient. Since we have control signals with $N_{c} = 20$ coefficients (10 for sinusoidal basis functions and 10 for cosinusoidal basis functions), from Eq.\eqref{d_limit} we obtain $\mu^{\rm dist.} \simeq 1.74 \cdot 10^{-8}$ eV. The periodic source $\Delta^{\rm dist.}_{j}(t)$ should have an energy amplitude less than $\mu^{\rm dist.}$ to not invalidate gate implementations.
Following a similar reasoning, we can define a bound $\mu^{\rm noise}$ on the energy amplitude of the withe noise $\Delta^{\rm noise}_{j}(t)$.
As before, we find  $
\mu^{\rm noise} = \sigma_{\Delta^{\rm noise}} \cdot \sqrt{2/\pi}. 
$
The average tolerated $\sigma^{\rm tol.}_{\Delta^{\rm noise}}$ that we obtained for 30 implementations of a CNOT gate is $\sigma^{\rm tol.}_{\Delta^{\rm noise}} \simeq 1.52 \cdot 10^{-8}$ eV. Hence 
we obtain that $\mu^{\rm noise} \simeq 1.21 \cdot 10^{-8}$ eV.

The comparison of $\mu^{\rm noise}$ and $\mu^{\rm dist.}$ suggests that our optimal control solutions may tolerate a little better the distortion that affect their spectra than the white noise. This result holds for all gates of the simulated universal set as is possible to observe comparing the two disturb sources using the data from Table \ref{table:noise} and Table \ref{table:dist}.

As is possible to observe from Table \ref{table:inf}, the obtained infidelities when no disturbances are superimposed on control signals are slightly below the target threshold. As a consequence, the optimal control solutions disturbed by noise and spectral distortion tend to easily exceed the infidelity threshold.
The aforementioned consideration and the choice to set the same threshold for the ideal and the disturbed gate implementations probably lead to tolerable noise and spectral distortion levels smaller than necessary.
Allowing a tolerable error on the infidelity of disturbed gate implementations could help to define less strict tolerable noise and spectral distortion levels.

\section{Conclusions}\label{s:conclu}

In this letter we have benchmarked the robustness in the implementation of quantum gates in a two-transmon system,
where each transmon is approximated as a three-level system with a single non-computational state. 
Optimal control signals to implement different gates from a universal set were obtained using the CRAB 
algorithm, and different sources of electronic imperfections were modeled as signal noise and 
spectral distortions. 
We report that we have been able to implement all gates from the chosen universal set with an infidelity below $10^{-2}$ and with a fixed control time $T=40$ ns, exploiting the CRAB optimization method. We also note that all optimization attempts have been able to define the control solutions at the first try, in spite of the errors caused by non-computational energy levels. 

We observed that to obtain a proper gate implementation it is mandatory to minimize noise and spectral distortion in control signals. 
We found the introduction of noise in control signals seems to be slightly less tolerated than spectral distortion. In future studies, it may be interesting to test the resilience of the optimal control solutions against colored noises and spectral distortion with different weight for each basis coefficient.
The numerical analysis allowed us to define empirical energy bounds for the Gaussian white noise (i.e. $\sigma_{\Delta^{\rm noise}}$) and the spectral distortion (i.e. $\sigma_{\Delta}$) that can be tolerated by optimal control solutions \cite{kallush2014quantum,lloyd2014information,muller2020information}. 

We also observed that some optimal control signals showed a significantly higher tolerance to disturbances.  For some gates we observed $\sim 3$ dB between $\sigma_{\Delta^{\rm noise}}$ of the average and the best performing gate implementations with noisy signals (see Table \ref{table:noise}). Similarly, for perturbations in the coefficients we observed $\sim 6$ dB between $\sigma_{\Delta}$ of the average and the best performing gate implementations (see Table \ref{table:dist}). 
It could be interesting to study whether the optimal control solutions that tolerated higher disturbances share some common characteristics. In case their best behavior could be treated deterministically, it may be possible to evolve CRAB to directly synthesize control solutions, capable of sustaining the typical disturbances of a target experimental system.

\bibliographystyle{elsarticle-num}
\bibliography{./bibliography}
\end{document}